\documentclass{moriond}

\def\PLB{{\em Phys. Lett.}  B}
\def\PRL{\em Phys. Rev. Lett.}

\def\be{\begin{equation}}
\def\ee{\end{equation}}
\def\bea{\begin{eqnarray}}
\def\eea{\end{eqnarray}}

\newcommand{\Photo}{\vspace{0.5cm}\includegraphics[height=35mm]{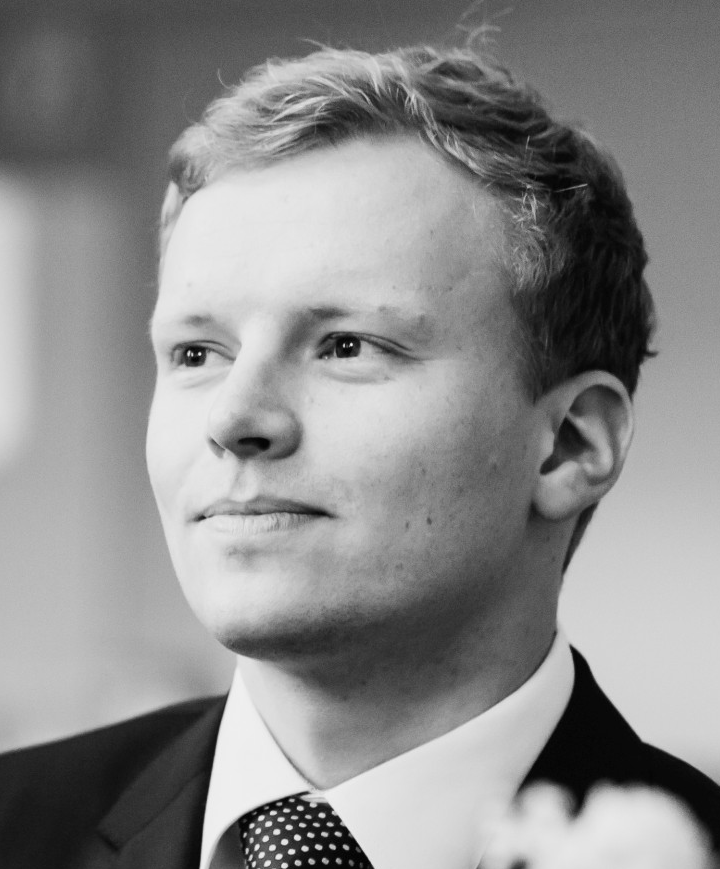}}

\begin{document}
\vspace*{4cm}
\title{HIGH-$p_T$ MULTI-JET FINAL STATES AT ATLAS AND CMS}

\author{ CLEMENS LANGE on behalf of the ATLAS and CMS collaborations }

\address{Physics Institute, University of Zurich, Winterthurerstrasse 190, 8057 Zurich, Switzerland}

\maketitle\abstracts{
The increase of the centre-of-mass energy of the Large Hadron Collider (LHC) to 13 TeV has opened up a new energy regime. Final states including high-momentum multi-jet signatures often dominate beyond standard model phenomena, in particular decay products of new heavy particles. While the potential di-photon resonance currently receives a lot of attention, multi-jet final states pose strong constraints on what physics model an observation could actually be described with. In this presentation, the latest results of the ATLAS and CMS collaborations in high transverse momentum multi-jet final states are summarised. This includes searches for heavy resonances and new phenomena in the di-jet mass spectrum, di-jet angular distributions, and the sum of transverse momenta in different event topologies. Furthermore, results on leptoquark pair production will be shown. A particular focus is laid on the different background estimation methods.}

\section{Introduction}

The increased centre-of-mass energy of the Large Hadron Collider (LHC) has opened up a new energy regime. In order to compare the potential to create new heavy particles in the LHC's proton-proton collisions at the new centre-of-mass energy of 13~TeV (Run~2) and the previously reached 8~TeV (Run~1), one needs to relate the parton luminosities. An integrated luminosity of 3~fb$^{-1}$ was collected in 2015 at $\sqrt{s} = 13$~TeV by the ATLAS~\cite{Aad:2008zzm} and CMS~\cite{Chatrchyan:2008aa} experiments, while the one Run~1 dataset at 8~TeV amounts to about 20~fb$^{-1}$. For masses of about 1~TeV, the luminosity ratio of 13 over 8~TeV is about 3 for the sum of quark-antiquark annihilation and about 6 for gluon-gluon fusion. This means that the experiments' sensitivities with the Run~2 dataset to new heavy particles approximately matches the one of Run~1 for particles with masses of about 1~TeV. For heavier particles, the experiments' reach with the available data already surpasses the sensitivity of the 8~TeV dataset.

Heavy new particles are predicted by a large number of models that try to explain phenomena that are not described by the standard model (SM). Final states including partons hereby often dominate these phenomena beyond the standard model (BSM). In the particle detectors, these are observed as multi-jet final states, which are the focus of this discussion.

\subsection{Relation of multi-jet final states to a potential di-photon resonance}\label{subsec:diphoton}

The small bump found in the di-photon spectrum around an invariant mass of 750~GeV in both the ATLAS and CMS experiments in 2015 data~\cite{Aaboud:2016tru,Khachatryan:2016hje} has caused a lot of excitement in the particle physics community. Should it turn out that this is a real new particle, one needs to consider how it relates to other final states. A neutral resonance cannot directly couple to photons---it needs a loop of charged particles such as a top quark or a W boson in decay and production. This implies that there should be more than just a di-photon resonance. The searches presented in the following constrain what physics model this potential resonance could actually be described with.

\subsection{Differentiating multi-jet final states with and without leptons}\label{subsec:finalstates}

When reviewing searches for new physics signals in multi-jet final states, one can observe a major difference between pure multi-jet final states and final states with leptons with respect to the background estimation methods employed. Final states without leptons are dominated by QCD multi-jet events. For these, Monte Carlo simulated event samples often do not have sufficient statistics. Furthermore, the different contributions to the background sample, which are partly instrumental, are difficult to model. Therefore, mostly functional forms are used for the background estimation.

For leptons plus jets final states, the multi-jet background is significantly reduced by requiring the presence of one or more lepton(s). This significantly changes the composition of the background processes, where now not necessarily only one is dominant anymore. This allows---and to some extent also requires---to estimate the background components individually.

\section{0-lepton and multi-jet background dominated final states}

\begin{figure}
\begin{minipage}{0.98\linewidth}
\centerline{\includegraphics[width=0.99\linewidth]{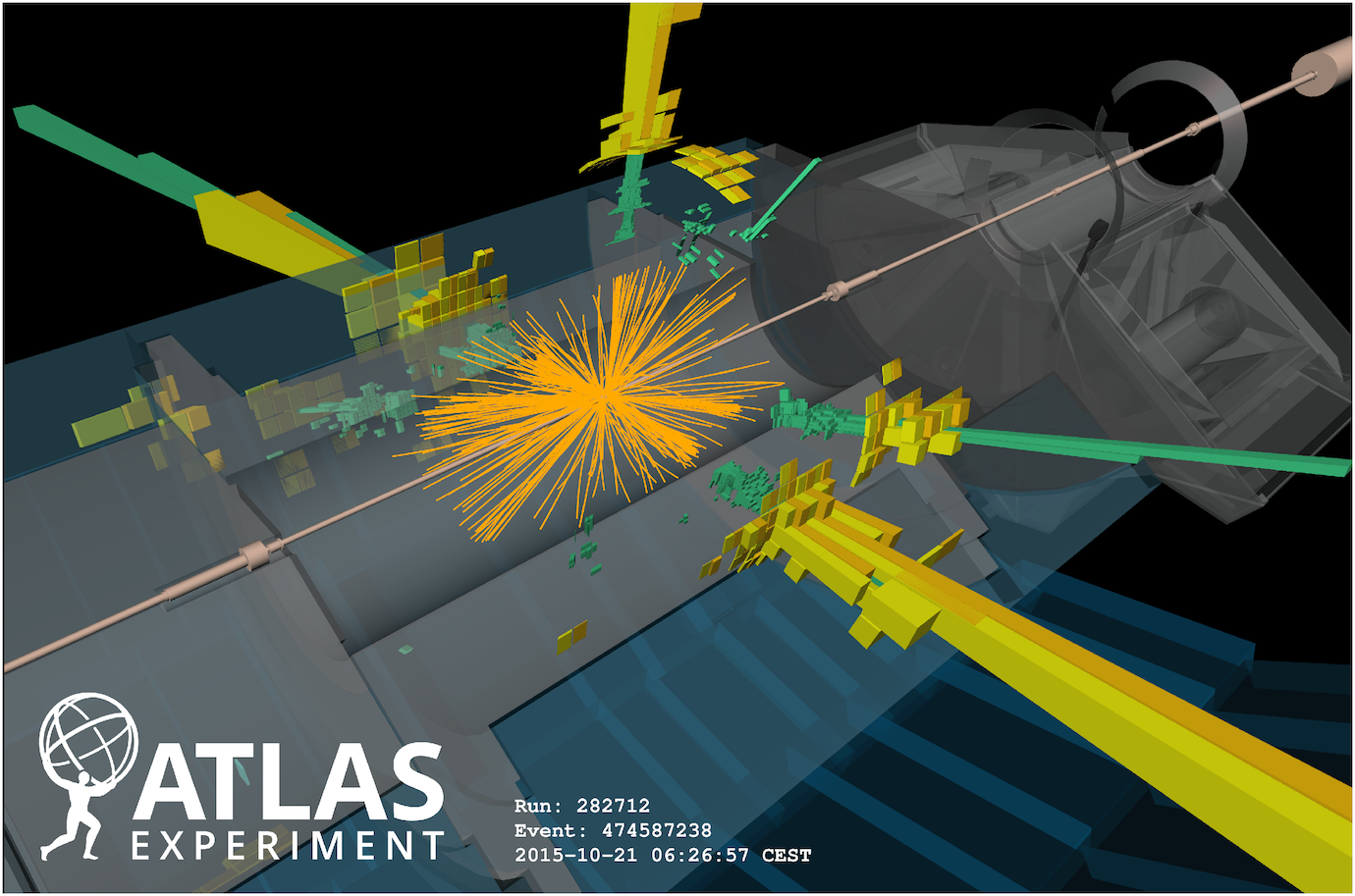}}
\end{minipage}
\caption[]{Event display of a multi-jet event in the ATLAS detector~\cite{Aad:2015mzg}.}
\label{fig:evdisplay}
\end{figure}

Final states, for which no attempt is made to reconstruct leptons are dominated by multi-jet events. The searches performed in these final states that are presented here take into account two to more than ten objects in the final state. An example event display of a multi-jet event is shown in Fig.~\ref{fig:evdisplay}.

\subsection{Background estimation}

As described above, in pure multi-jet final states, the background estimation is usually performed using functional shapes. The challenge hereby lies in the choice of a function that is able to describe the background while maintaining sensitivity of the analysis to new physics signals. The initial choice of the background functional form is mostly arbitrary with the assumption of a falling spectrum in the variable of interest such as the di-jet invariant mass spectrum. It is therefore mandatory to define a procedure to choose the functional form and also the number of parameters based on statistical tests. An example of a background function as used in Ref.~\cite{Khachatryan:2015dcf} is given in Eq.~\ref{eq:dijet}:
\begin{equation}
\frac{d\sigma}{dm_{jj}} = \frac{P_0 (1-x)^{P_1}}{x^{P_2+P_3 \ln(x)}},
\label{eq:dijet}
\end{equation}
where $P_0$ to $P_3$ are the potential free parameters and $x = m_{jj} / \sqrt{s}$. To decide on the number of parameters to be used eventually, in this example fits to the data spectrum are performed starting with only two free parameters. Then a fit with three parameters is performed and an F-test is used to decide whether the higher number of parameters improves the modelling of the data at a certain confidence level. Should this be the case, a fit with four free parameters is performed and again the F-test is used, otherwise one sticks with the fit function with the lower number of free parameters.

In other analyses presented here, different approaches are taken, but they are all based on extensive statistical tests. MC-simulated events are used for validation. In case a bias is observed, this needs to be taken into account in the analysis, e.g.\ by changing the background estimation method or correcting for the bias.

\subsection{Di-jet resonance and angular analyses}

The topologically most simple analyses are the di-jet resonance searches~\cite{Khachatryan:2015dcf,ATLAS:2015nsi} and the ones looking for new phenomena in angular distributions~\cite{ATLAS:2015nsi,CMS:2015djr}. To select the data, both ATLAS and CMS use single jet triggers with thresholds of 360 and 500~GeV, respectively. CMS additionally makes use of a trigger based on the scalar transverse momentum sum of all objects in the event ($H_T$) with a threshold of 800~GeV. Events triggered like that are then required to contain at least two jets with transverse momenta ($p_T$) greater than 440 and 50~GeV, respectively. In the ATLAS analyses, they both need to be in the centre of the detector within absolute rapidity $|y^{\star}| < 1.7$. The CMS analyses have a symmetric $p_T$-cut of 30~GeV for both jets, but additionally apply a cut on the invariant di-jet mass $m_{jj} > 1.2$~TeV and the pseudorapidity difference between the two jets $|\Delta \eta_{jj}| < 1.3$. With these cuts applied, the triggers are found to be almost 100\% efficient. The discriminants used in the analyses are the di-jet invariant mass, $m_{jj}$, for the di-jet resonance searches and the rapidity difference as defined in Eq.~\ref{eq:angular}:
\begin{equation}
\chi = e^{2|y^{\star}|} \sim \frac{1 + \cos \theta^{\star}}{1 - \cos \theta^{\star}},
\label{eq:angular}
\end{equation}
where $\theta^{\star}$ is defined as the polar angle in the di-jet centre-of-mass frame, for the di-jet angular analyses.

These distributions are sensitive to a large number of new physics model signatures such as quantum black holes, excited quarks, W'/Z' particles, and contact interactions; for review see Refs.~\cite{Boelaert:2009jm,Harris:2011bh}. However, the analyses aim to set model-independent limits to make them easily accessible to the theory community.

\begin{figure}
\begin{minipage}{0.48\linewidth}
\centerline{\includegraphics[width=0.97\linewidth]{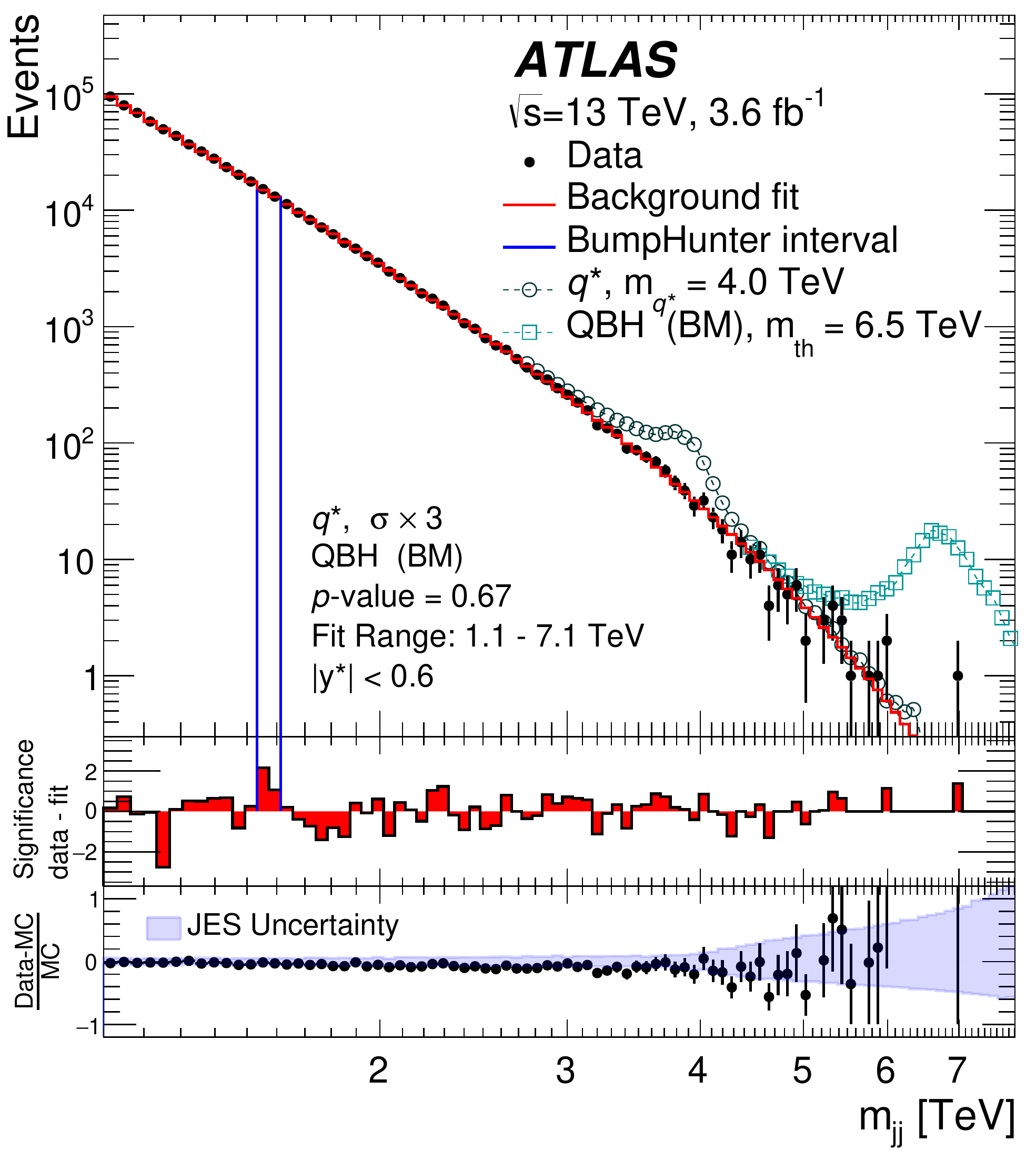}}
\end{minipage}
\hfill
\begin{minipage}{0.48\linewidth}
\centerline{\includegraphics[width=0.99\linewidth]{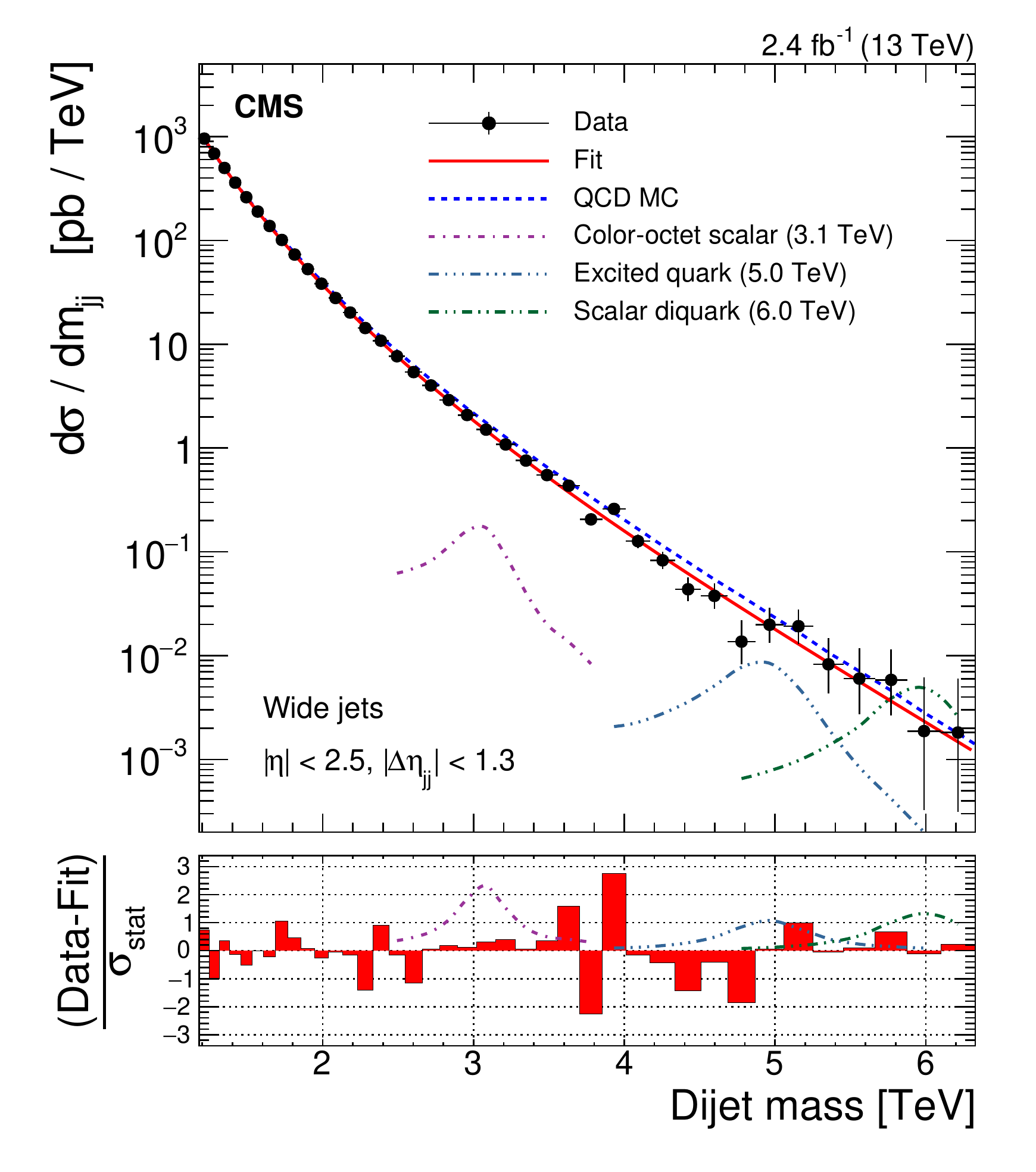}}
\end{minipage}
\caption[]{The reconstructed di-jet mass distribution for (left) ATLAS~\cite{ATLAS:2015nsi} and (right) CMS~\cite{Khachatryan:2015dcf}. In both figures potential new physics signals are also shown.}
\label{fig:dijet}
\end{figure}

The di-jet resonance spectra for both the ATLAS and CMS analyses are shown in Fig.~\ref{fig:dijet}. No signal is found, thus limits are set, which both analyses significantly extend with respect to the 8~TeV results.

\begin{figure}
\begin{minipage}{0.48\linewidth}
\centerline{\includegraphics[width=0.95\linewidth]{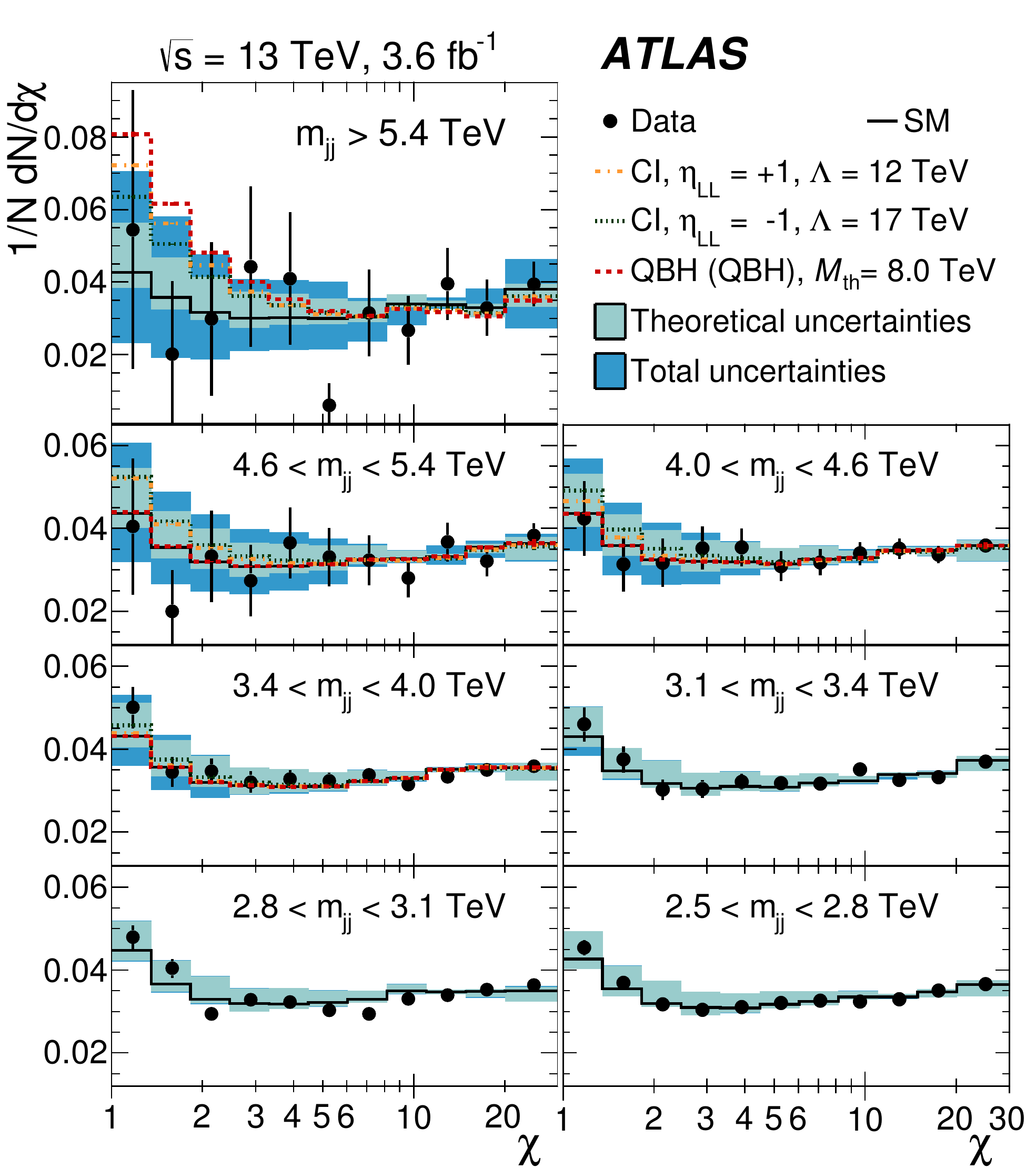}}
\end{minipage}
\hfill
\begin{minipage}{0.48\linewidth}
\centerline{\includegraphics[width=0.99\linewidth]{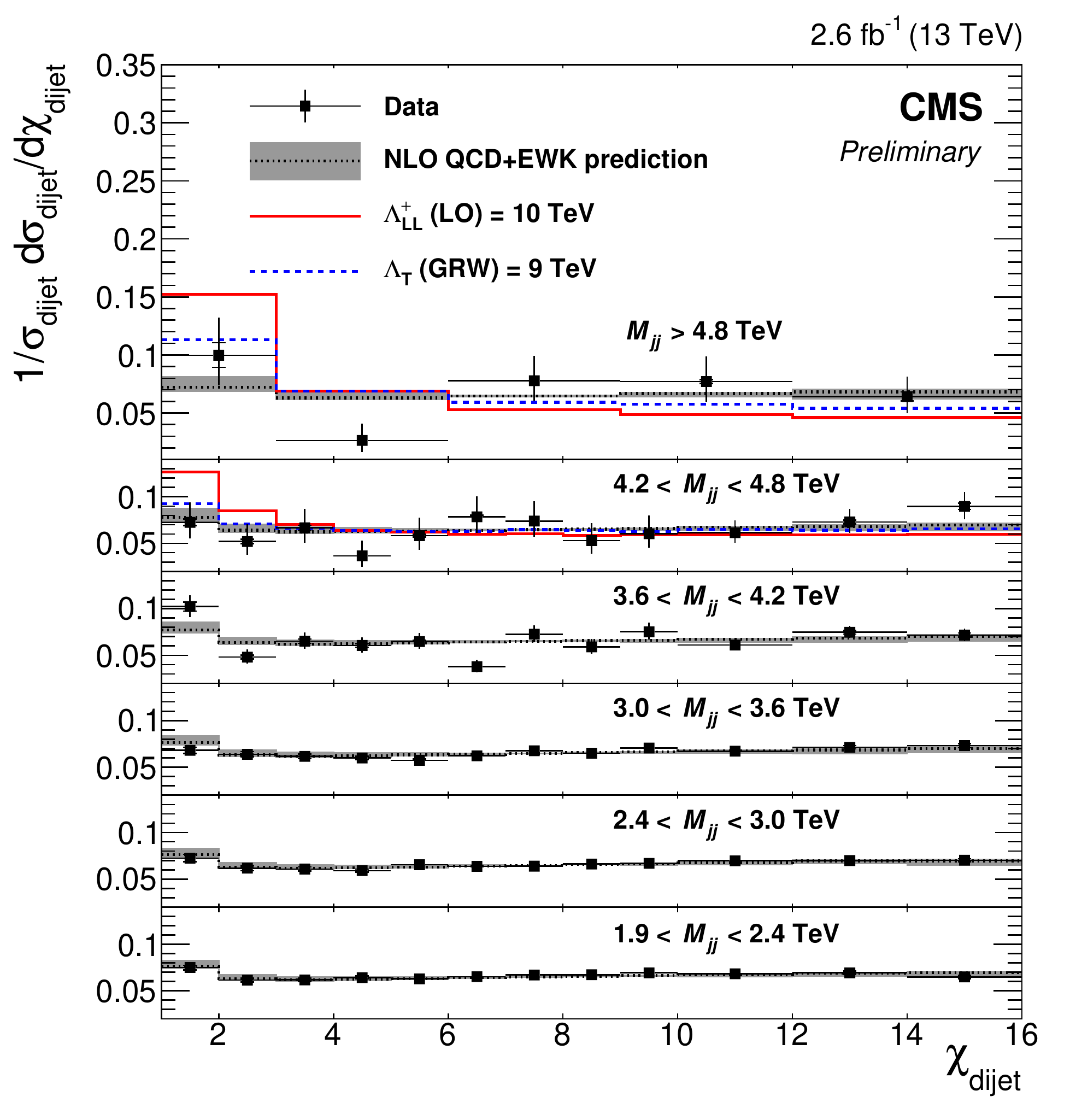}}
\end{minipage}
\caption[]{The reconstructed di-jet rapidity difference compared to next-to-leading order theory predictions including electroweak corrections in different di-jet invariant mass windows for (left) ATLAS~\cite{ATLAS:2015nsi} and (right) CMS~\cite{CMS:2015djr}. Expected distributions for new physics signals are also shown.}
\label{fig:angular}
\end{figure}

The di-jet angular distributions are shown in Fig.~\ref{fig:angular}. The distribution is expected to be uniform for Rutherford scattering. Higher order corrections (here next-to-leading order theory predictions including electroweak corrections) change the distributions significantly underlining their importance, since new physics signals also predict deviations, which are even larger. No significant deviations from the theory prediction are observed.

\subsection{Search for di-jet resonances with one or two jets identified as b-jets}

\begin{figure}
\begin{minipage}{0.48\linewidth}
\centerline{\includegraphics[width=0.99\linewidth]{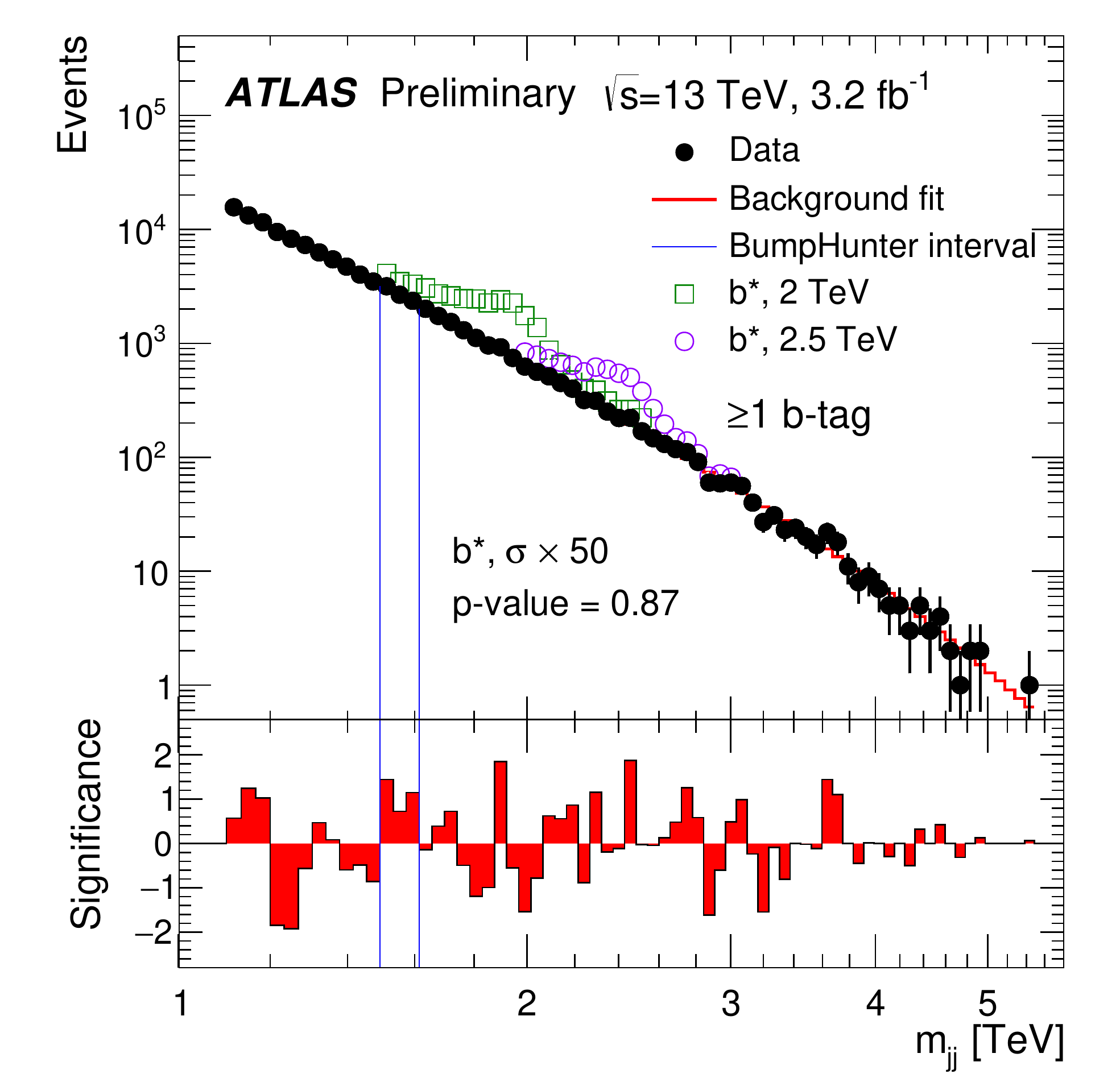}}
\end{minipage}
\hfill
\begin{minipage}{0.48\linewidth}
\centerline{\includegraphics[width=0.99\linewidth]{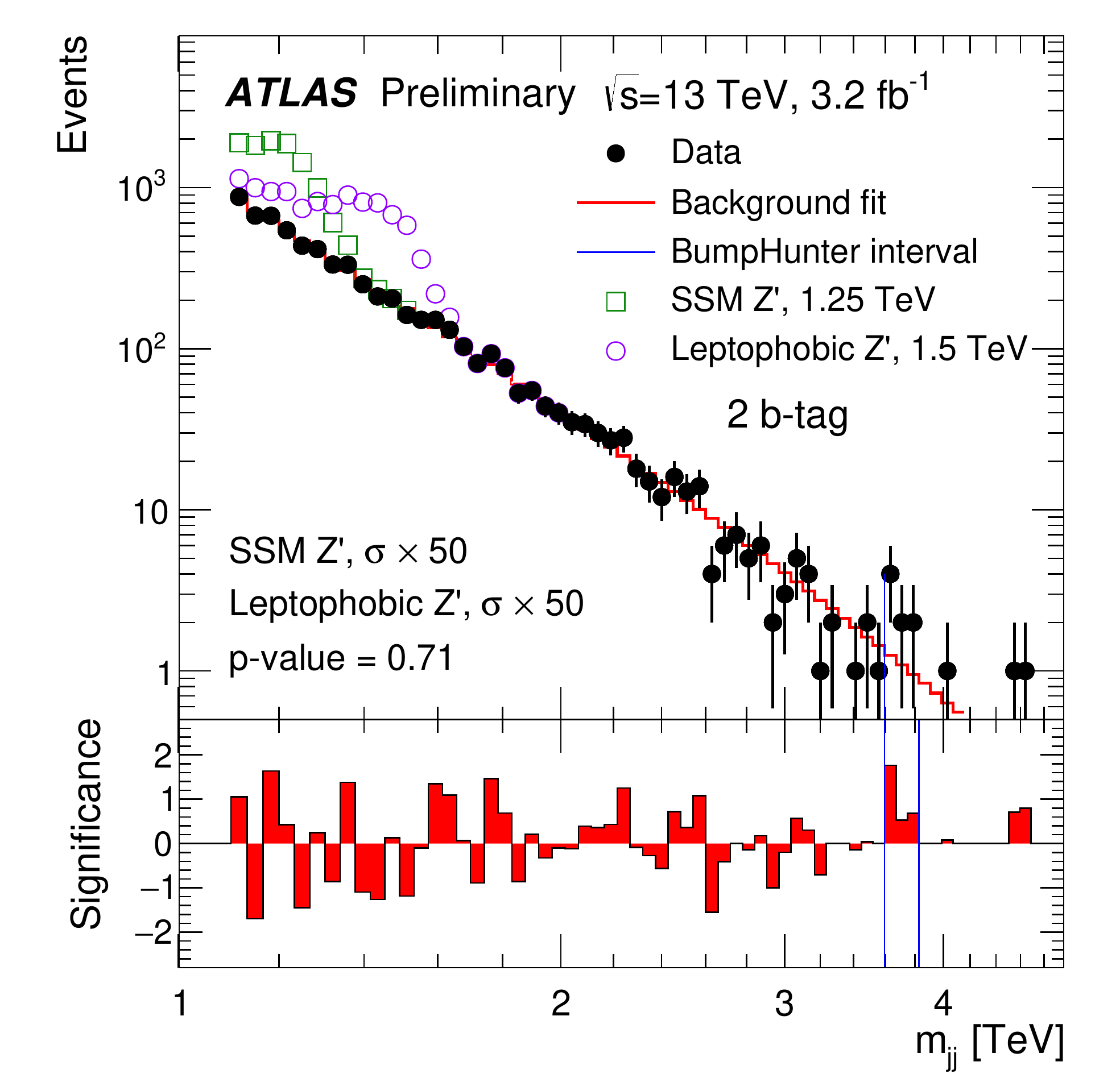}}
\end{minipage}
\caption[]{The reconstructed di-jet invariant mass distribution for (left) at least one identified b jet and (right) two identified b jets~\cite{Aaboud:2016nbq}. Expected distributions for new physics signals are also shown.}
\label{fig:bjet}
\end{figure}

The di-jet resonance search has recently been extended to events, in which one or both of the two jets are identified to contain B hadrons~\cite{Aaboud:2016nbq}. This analysis, performed by the ATLAS collaboration, uses the same background estimation method as the di-jet analysis discussed above. However, the single and double b-tag categories are sensitive to different signal hypotheses such as 4$^\mathrm{th}$ generation b-quark models~\cite{Baur:1987ga,Baur:1989kv} and Z' models~\cite{Langacker:2008yv,Chiang:2014yva}. A 4$^\mathrm{th}$ generation b quark is excluded in the range of 1.1--2.1~TeV while the analysis is not yet sensitive to sequential SM Z' particles. The di-jet invariant mass spectra are shown in Fig.~\ref{fig:bjet}.

\subsection{Search for new physics in multi-jet spectra}

Taking more than two jets into account, the analyses become particularly sensitive to signs of strong gravity and black holes in general (see e.g. Ref.~\cite{ArkaniHamed:1998rs}). The ATLAS collaboration has performed a search in 3--8 jet final states, looking for signs of thermal black holes in the $H_T$ spectrum~\cite{Aad:2015mzg}. The background estimation strategy is hereby such that fits of ten different functions are performed to low values of $H_T$, which are then validated at medium $H_T$ and one of the functions chosen for the final analysis. This decision is based on a bootstrap method: incremental datasets of integrated luminosities of 6.5~pb$^{-1}$, 74~pb$^{-1}$, 440~pb$^{-1}$ and 3.0~fb$^{-1}$ are used to define the control regions and to confirm the background estimation method. No signal is found and limits are set.

This kind of analysis is extended further by the CMS collaboration taking into account electrons, photons, and muons in addition to jets~\cite{CMS:2015iwr}. The search is performed in the spectrum of the scalar sum of the objects' transverse momenta, the so-called $S_T$ distribution. The background is normalised by a fit in the low $S_T$ region in each of the different object multiplicity bins. Two to more than ten objects are considered in the analysis, but again no signal is found. The limits are significantly expanded with respect to the results at $\sqrt{s} = 8$~TeV.

\section{Jets + leptons final states}

As described above, the jets + leptons final states follow a different analysis strategy. While the analyses discussed in the previous section rely on jet triggers, the analyses presented in the following are based on lepton triggers. The requirement of an isolated lepton (here only muons and electrons are considered) drastically suppresses the multi-jet background and allows for different background estimation methods.

\subsection{Search for strong gravity}

\begin{figure}
\begin{minipage}{0.48\linewidth}
\centerline{\includegraphics[width=0.99\linewidth]{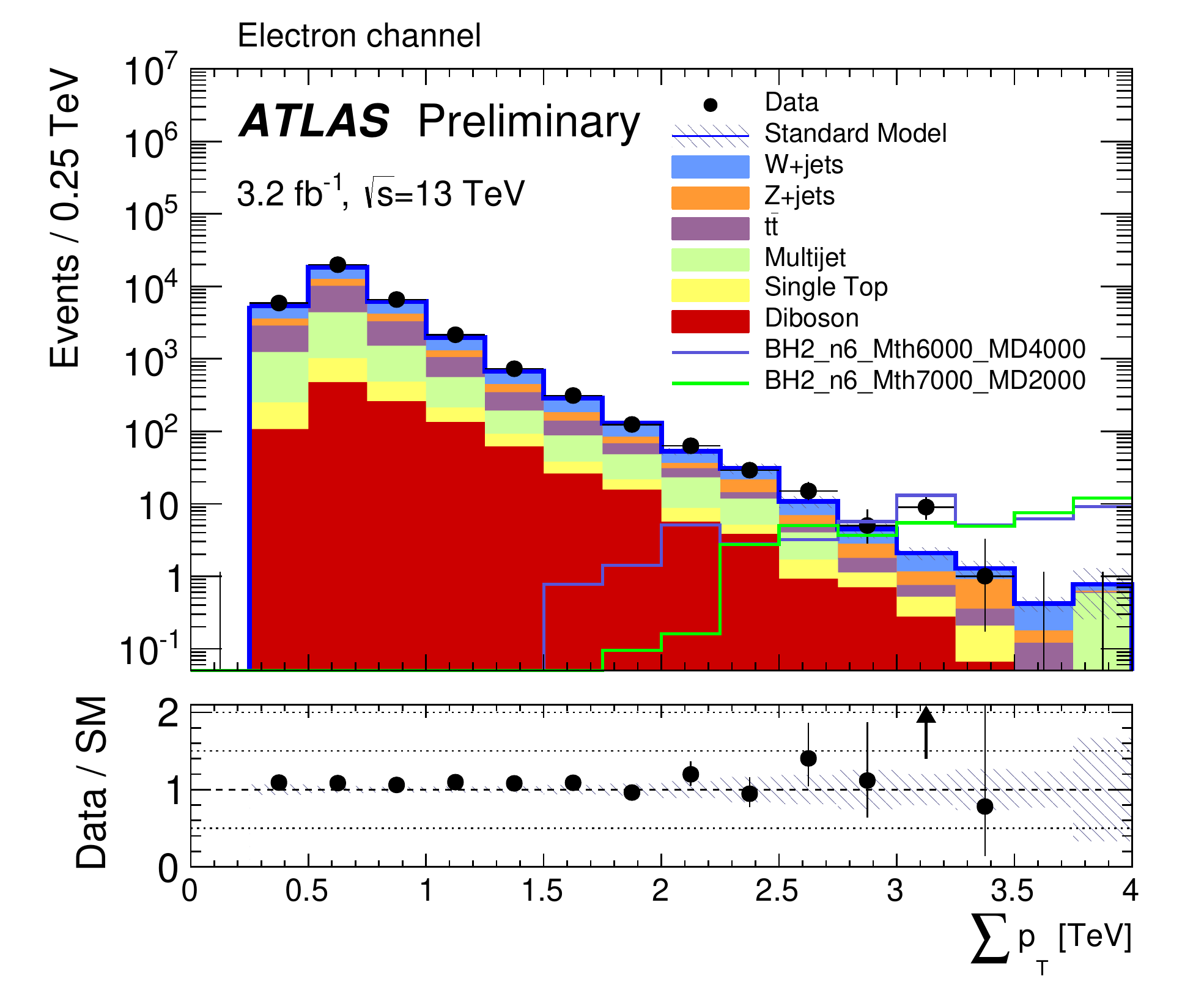}}
\end{minipage}
\hfill
\begin{minipage}{0.48\linewidth}
\centerline{\includegraphics[width=0.99\linewidth]{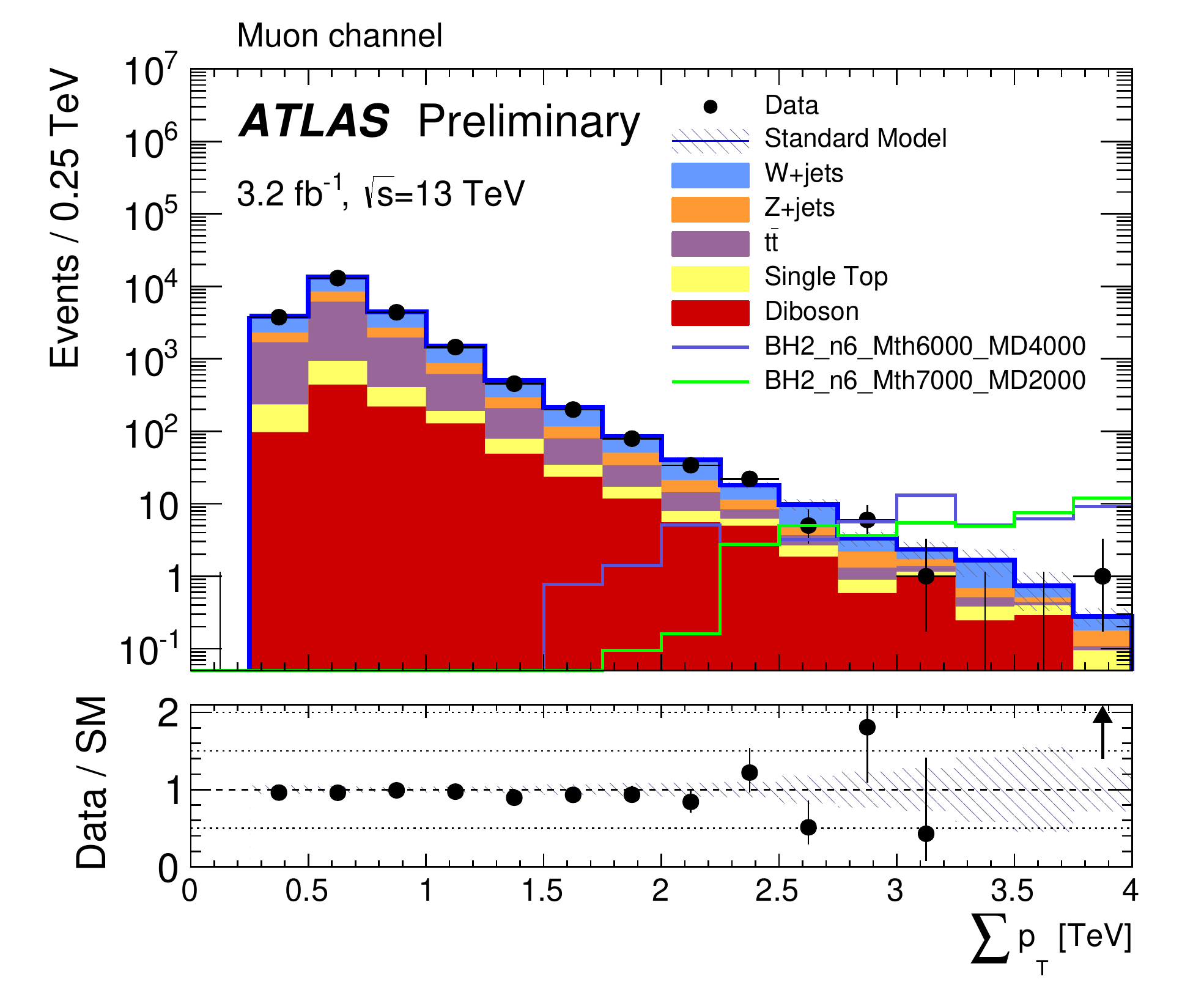}}
\end{minipage}
\caption[]{The reconstructed scalar $p_T$-sum distribution for (left) the electron and (right) the muon channel of the ATLAS search for strong gravity in lepton + jets events~\cite{ATLAS:2016006}. Expected distributions for different black hole models are also shown.}
\label{fig:strgrav}
\end{figure}

Black holes~\cite{ArkaniHamed:1998rs} and string balls~\cite{Dimopoulos:2001qe} are expected to decay democratically according to the degrees of freedom of the SM due to
the energy-momentum tensor coupling. This means that the signal strength of these hypothetical processes can be increased significantly by selecting a lepton. In the analysis performed by the ATLAS collaboration at least three high-$p_T$ objects (jets, muons or electrons) are selected and the search is conducted in the scalar $p_T$-sum of those objects~\cite{ATLAS:2016006}. The background modelling, which is taken from MC simulation, is confirmed in dedicated control regions and at low sum-$p_T$. Eventually, a combined fit in all control regions and the signal region is performed, the result of which is shown in Fig.~\ref{fig:strgrav}. No signal is found and exclusion limits are improved by 2--3~TeV depending on the black hole model parameters used.

\subsection{Search for pair production of second generation leptoquarks}

\begin{figure}
\begin{minipage}{0.48\linewidth}
\centerline{\includegraphics[width=0.99\linewidth]{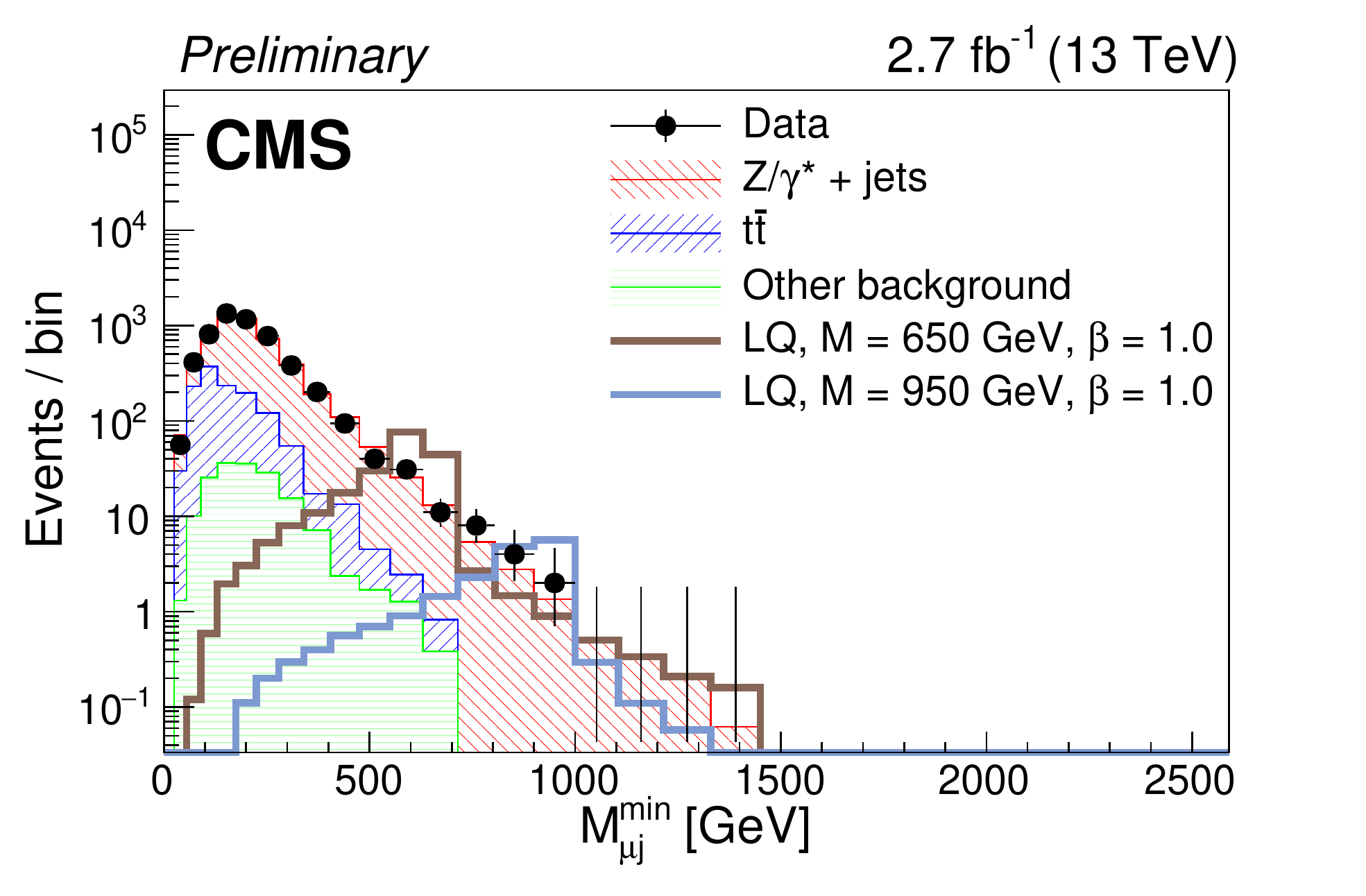}}
\end{minipage}
\hfill
\begin{minipage}{0.48\linewidth}
\centerline{\includegraphics[width=0.99\linewidth]{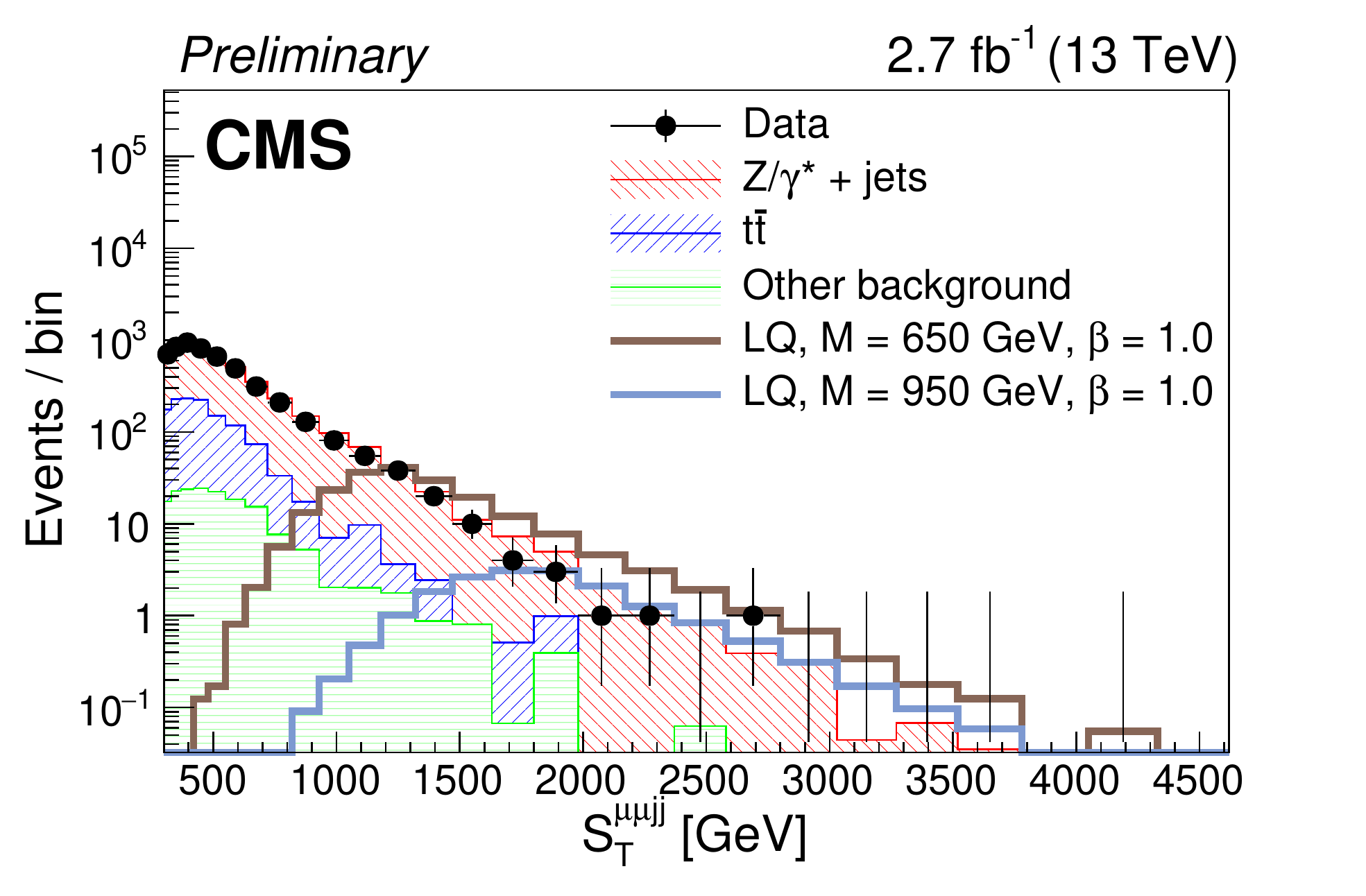}}
\end{minipage}
\caption[]{The reconstructed distributions of the (left) $p_T$-sum of the muon and jet pairs ($S_T$) and (right) the minimum of the muon-jet invariant mass of the CMS leptoquark search~\cite{CMS:2016qhm}. Expected distributions for the signal mass hypotheses of 650 and 900~GeV are also shown.}
\label{fig:lq}
\end{figure}

Leptoquarks are hypothetical particles that carry both baryon and lepton numbers. At the LHC, they are expected to be produced in pairs. The CMS collaboration has performed a search for these particles in the 2 muon + 2 jets final state~\cite{CMS:2016qhm}. The dominant background in this analysis is Drell-Yan production, which is suppressed by vetoing events for which the di-muon invariant mass is close to the Z boson mass. Furthermore, cuts are applied on the scalar $p_T$ sum of the muon and jet pairs ($S_T$), and the minimum of the muon-jet invariant mass. These cuts are optimised for each signal mass hypothesis ranging from 200--1500~GeV. The distributions for the 650 and 900~GeV signal mass hypotheses are shown in Fig.~\ref{fig:lq}. The dominant background processes (Drell-Yan and top quark pair production) are estimated in sidebands, and the final number of events is counted in each mass window. No sign of leptoquark production is found and production of leptoquarks with masses below 1150~GeV assuming a 100\% branching ratio to lepton + quark are excluded at 95\% confidence level.

\section{Conclusions and outlook}

\begin{figure}
\begin{minipage}{0.45\linewidth}
\centerline{\includegraphics[width=0.99\linewidth]{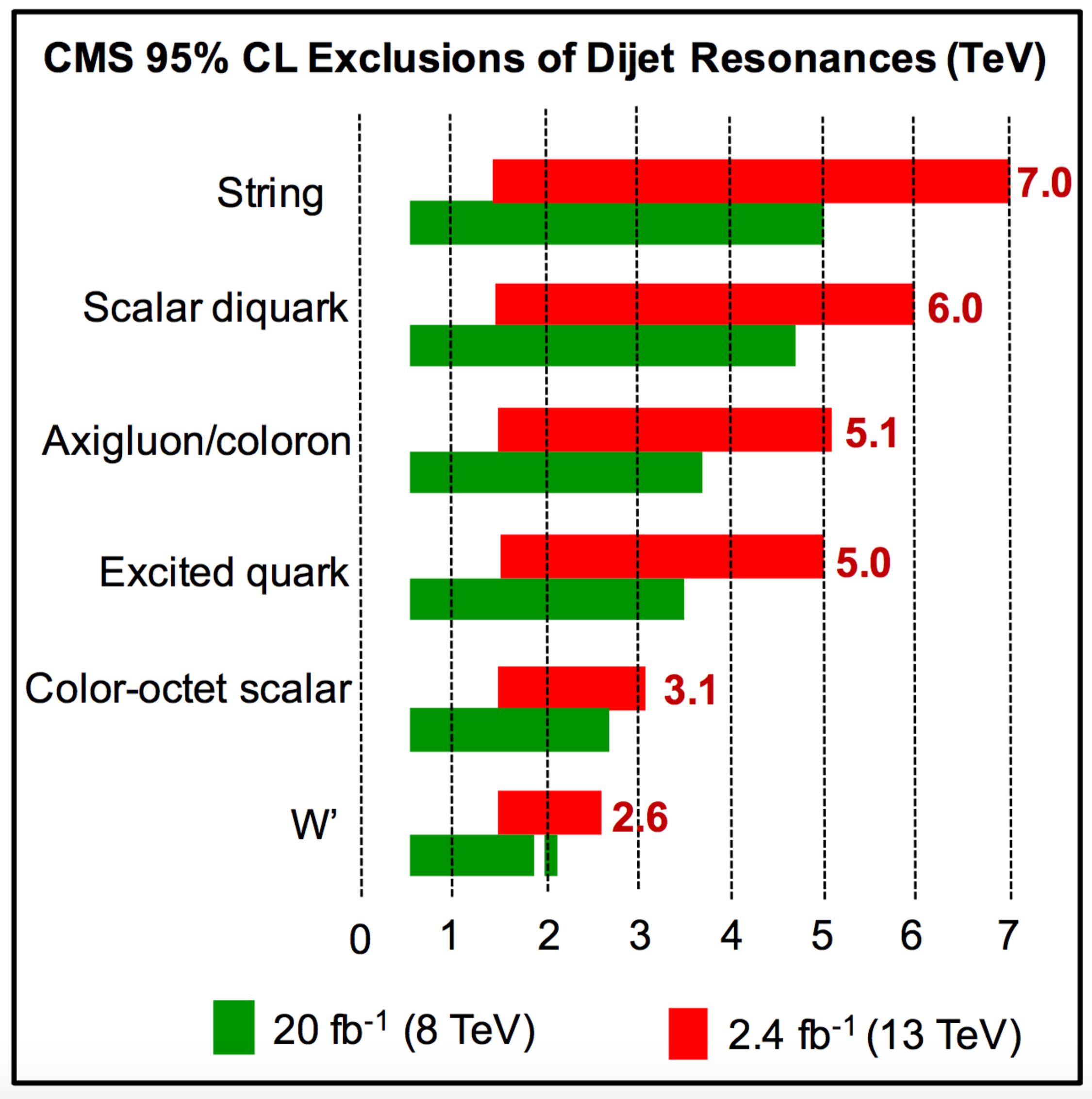}}
\end{minipage}
\hfill
\begin{minipage}{0.53\linewidth}
\centerline{\includegraphics[width=0.99\linewidth]{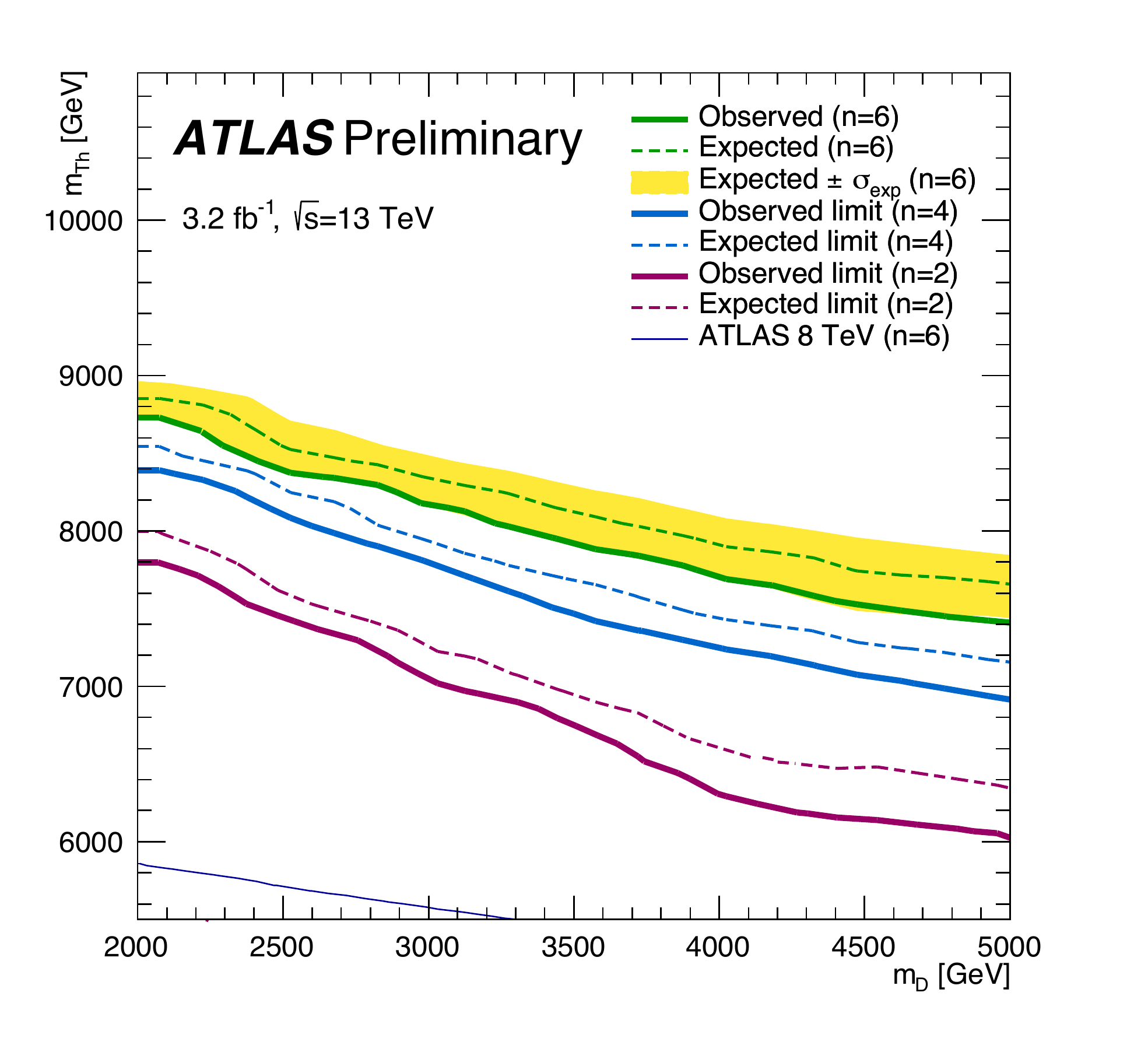}}
\end{minipage}
\caption[]{Left: Observed 95\% C.L.\ lower limits on di-jet resonance mass for the listed models obtained by the CMS experiment in Run~1 (green) and Run~2 (red) analyses~\cite{Khachatryan:2015dcf}. Right: Exclusion contours in the threshold mass ($m_\mathrm{Th}$), gravity scale $(m_\mathrm{D}$) plane for rotating black hole models~\cite{ATLAS:2016006}. Compare the solid green line (13~TeV result) to the thin blue line (8~TeV result) for six extra dimensions.}
\label{fig:comp}
\end{figure}

No matter if the di-photon bump is real or not, there are loads of reasons to search for new physics in multi-jet final states. The searches in these final states with the first collision data recorded at a centre-of-mass energy of 13~TeV have shown no signs of deviations from the SM expectation. However, largely thanks to parton luminosity scaling, the previously obtained limits have been significantly improved using only a fraction of the data statistics compared to the 8 TeV run. Two examples of how much the limits improved using 13~TeV data are shown in Fig.~\ref{fig:comp}. 2016, commencing the luminosity ramp-up at 13 TeV, will be another exciting year at the LHC.

\end{document}